\documentclass[a4paper,fleqn,usenatbib]{mnras}
\usepackage{newtxtext,newtxmath}
\usepackage[T1]{fontenc}
\usepackage{ae,aecompl}
\usepackage{graphicx}	% Including figure files
\usepackage{amsmath}	% Advanced maths commands
\usepackage{amssymb}	% Extra maths symbols

\usepackage{esvect}
\usepackage{color}
\usepackage{epstopdf}
\usepackage{ulem}

\newcommand*\mean[1]{\overline{#1}}

\begin{document}

% Title of the paper, and the short title which is used in the headers.
% Keep the title short and informative.
\title[Growth rate]{An estimation of the local growth rate from Cosmicflows-3 peculiar velocities}

% The list of authors, and the short list which is used in the headers.
% If you need two or more lines of authors, add an extra line using \newauthor
\author[ Dupuy et al.]{Alexandra Dupuy$^{1}$ \thanks{E-mail: dupuy@ipnl.in2p3.fr}, Helene M. Courtois$^{1}$, Bogna Kubik$^{1}$\\
% List of institutions
$^{1}$University of Lyon, UCB Lyon 1, CNRS/IN2P3, IPN Lyon, 69622 Villeurbanne, France\\
}

\maketitle
% Abstract of the paper
\begin{abstract}
This article explores three usual estimators, noted as $v_{12}$ of the pairwise velocity, $\psi_1$ and $\psi_2$ of the observed two-point galaxy peculiar velocity correlation functions.  These estimators are tested on mock samples of {\it Cosmicflows-3} dataset  \citep{Tully:2016aa} , derived from a numerical cosmological simulation, and also on a number of constrained realizations of  this dataset. Observational measurements errors and cosmic variance are taken into consideration in the study. The result is a local measurement of $f \sigma_8 = 0.43 \left( \pm 0.03 \right)_\mathrm{obs} \left( \pm 0.11 \right)_\mathrm{cosmic}$ out to $z=0.05$, in support of a $\Lambda$CDM cosmology. 
\end{abstract}

%%%%%%%%%%%%%%%%%%%%%introduction%%%%%%%%%%%%%%%%%%%%%%%%%%%%%

\section{Introduction}
\label{sec:intro}

Since the late 70's, several publications discussed the theory of galaxy pairwise peculiar velocity statistics, such as the 2-point peculiar velocity correlation function ($\psi_1$ and $\psi_2$) or the mean pairwise velocity \citep[$v_{12}$; ][]{Monin:1975aa,Davis:1977aa,Peebles:1980aa,Peebles:1987aa,Gorski:1988aa}. It has been shown that such statistics can be measured directly from only the radial part of peculiar velocities.  
Since these statistics are related to the growth factor of large scale structures $f=\Omega_m^\gamma$, where $\gamma$ is the growth index \citep{Lahav:1991aa}, observed peculiar velocities can be used as cosmological probes to estimate the matter density parameter $\Omega_m$ \citep{Ferreira:1999ab,Juszkiewicz:1999aa}. However, $f$ and $\sigma_8$, the amplitude of the density fluctuations on 8 Mpc $h^{-1}$ scales (where $h=H_0/100$ and $H_0$ is the Hubble constant), are degenerate and cannot be constrained separately when using only galaxy peculiar velocity data.

The first attempts of constraining cosmological parameters such as the density parameter have been made by \cite{Peebles:1976aa}, \cite{Kaiser:1990aa} and \cite{Hudson:1994aa}. Later, \cite{Juszkiewicz:2000aa} gave $\Omega_m = 0.35 \pm 0.15$ with measurements of the mean pairwise velocity on the {\it Mark III} catalog of radial peculiar velocities of roughly 3,000  spiral and elliptical galaxies \citep{Willick:1995aa,Willick:1996aa,Willick:1997aa}. Then, \cite{Feldman:2003aa} obtained a very similar value of $\Omega_m = 0.30 ^{+ 0.17} _{- 0.07}$ and also measured  $\sigma_8 = 1.13 ^{+0.22} _{- 0.23}$. This was done by using the same estimator of \cite{Juszkiewicz:2000aa} , but on a much larger dataset combining peculiar velocities of approximately 6,400 galaxies extracted from several catalogs: {\it Mark III}, {\it Spiral Field I-Band} \citep{Giovanelli:1994aa,Giovanelli:1997ab,Giovanelli:1997aa,Haynes:1999aa,Haynes:1999ab}, {\it Nearby Early-type Galaxies Survey} \citep{da-Costa:2000aa} and the {\it Revised Flat Galaxy Catalog} \citep{Karachentsev:2000aa}.

A decade later, datasets improved and methods of analyzing peculiar velocity to constrain cosmology evolved. Some authors proposed new statistical methodologies using observed peculiar velocities, which differs from $v_{12}$ and $\psi_{1,2}$, in order to constrain the growth rate of large scale structures and related parameters. On the one hand, \cite{Hudson:2012aa} measured $f\sigma_8 \equiv \Omega_m^{0.55} \sigma_8 = 0.40 \pm 0.07$ by comparing the observed peculiar velocities of 245 supernovae (extracted from a compilation dubbed the {\it First Amendment}, A1) to the galaxy density field predicted by the {\it Point Source Catalogue Redshift Survey} \citep[PSCz,][]{Saunders:2000aa}. This method has been later applied by \cite{Carrick:2015aa} on galaxies from the 2M++ redshift compilation \citep{Lavaux:2011aa}, finding a much more accurate estimate of the growth factor $f\sigma_8 = 0.401 \pm 0.024$. On the other hand, \cite{Johnson:2014aa} analyzed the two-point statistics of the peculiar velocity field and obtained $f\sigma_8 = 0.418 \pm 0.065$ from a sample gathering peculiar velocities of $\sim$ 9,200 galaxies from the {\it Six Degree Field Galaxy Survey} peculiar velocity catalog \citep[6dFGS,][]{Jones:2004aa,Jones:2006aa,Jones:2009aa} and various supernovae distance measurements. Alternatively, using again the same two-point statistic ($v_{12}$) as \cite{Juszkiewicz:2000aa} and \cite{Feldman:2003aa}, applied on the {\it Cosmicflows-2} catalog containing $\sim$8,000 galaxy distances \citep[CF2,][]{Tully:2013aa}, \cite{Ma:2015aa} found surprising results:  $\Omega_m^{0.6} h = 0.102 ^{+ 0.384} _{- 0.044}$ and $\sigma_8 = 0.39 ^{+0.73} _{- 0.1}$. Moreover, by measuring the covariance of radial peculiar velocities in two catalogs, a sample of 208 low redshift supernovae (named {\it SuperCal}) and a set of roughly 9,000 peculiar velocities from 6dFGS, \cite{Huterer:2016aa} evaluated $f \sigma_8 = 0.428 ^{+0.048} _{-0.045}$ at $z = 0.02$. Finally, \cite{Adams:2017aa} measured $f\sigma_8 = 0.424 ^{+0.067} _{-0.064}$ by modelling the cross-covariance of the galaxy overdensity and peculiar velocity fields and applying their analysis to the observed peculiar velocities from the 6dFGS data.

Despite the increase in number of measurements and in redshift coverage, peculiar velocity catalogs are still not large enough and remain noticeably sparse at large distances. Hence, growth rate estimations are affected by uncertainties introduced by cosmic variance as they are obtained from local observations. \cite{Hellwing:2016aa} discussed the effect of the observer location in the universe on the derivation of two-point peculiar velocity statistics ($v_{12}$, $\psi_1$ and $\psi_2$) by considering two sets of randomly chosen observers and Local Group -like observers. The authors showed that the local environment, especially the Virgo cluster, systematically introduces deviations from predictions. 

More recently, \cite{Nusser:2017aa} measured $f\sigma_8 = 0.40 \pm 0.08$ by measuring velocity - density correlations on the largest and most recent catalog of $\sim$ 18,000 accurate galaxy distances, {\it Cosmicflows-3} \citep[CF3,][]{Tully:2016aa}. And, last but not least, \cite{Wang:2018aa} analyzed the peculiar velocity correlation functions through the estimators $\psi_1$ and $\psi_2$ applied to the {\it Cosmicflows} catalogs (CF2 and CF3) to constrain cosmological parameters: $\Omega_m = 0.315^{+0.205}_{-0.135}$ and $\sigma_8 = 0.92^{+0.440}_{-0.295}$.

On the grounds of the previous literature works introduced above, this article studies two classical two-point peculiar velocity statistics, using radial peculiar velocities provided by the {\it Cosmicflows-3} catalog, to constrain the local value of the growth rate factor $f\sigma_8$. Its structure is organized as follows. Section 2 provides details on the peculiar velocity data used for the analysis. The methodology of two-point correlation functions of peculiar velocities is described in Section 3 and is tested and validated on mocks in Section 4. Section 5 shows the velocity statistics measured on observed peculiar velocities. The main result of this article, the estimate of the growth rate from Cosmicflows radial peculiar velocities, is discussed in Section 6.

%%%%%%%%%%%%%%%%%%%%%data%%%%%%%%%%%%%%%%%%%%%%%%%%%%%

\section{Data}
\label{sec:data}

\subsection{Observed peculiar velocities: Cosmicflows-3}

%The CF2 catalog \citep{Tully:2013aa} provides accurate distances for 8,188 galaxies which can be redistributed within 4,899 groups, with a homogeneous volume coverage up to 80 Mpc $h^{-1}$. Most of these distances are derived with the Tully-Fisher (TF) relation \citep{Tully:1977aa}, linking the luminosity to the HI line width for spiral galaxies, and the Fundamental Plane (FP) relation \citep{Djorgovski:1987aa,Dressler:1987aa} for elliptical galaxies. Other distance measurements are obtained from various methods where applicable such as Cepheids, Tip of the Red Giant Branch, type Ia Supernovae, and surface brightness fluctuations.

The latest CF3 catalog \citep{Tully:2016aa} provides distances for 17,648 galaxies which can be redistributed within 11,936 groups, up to 150 Mpc $h^{-1}$. It is an expansion of the previous CF2 catalog \citep{Tully:2013aa}. It contains 8,188 galaxy distances with a homogeneous volume coverage up to 80 Mpc $h^{-1}$, mostly derived with the Tully-Fisher (TF) relation \citep{Tully:1977aa}, linking the luminosity to the HI line width for spiral galaxies, and the Fundamental Plane (FP) relation \citep{Djorgovski:1987aa,Dressler:1987aa} for elliptical galaxies. The main additions to the CF3 catalog are new distances, obtained with the FP relation from 6dFGS, and distances computed with the TF relation. About 60 percent of CF3 distances are therefore measured with the FP method and mostly located in the Southern celestial hemisphere, while around 40 percent of distances are obtained with the TF relation. Few distance measurements are obtained from various methods where applicable such as Cepheids, Tip of the Red Giant Branch, type Ia Supernovae, and surface brightness fluctuations.

From the distance $d$ of a galaxy and its redshift $z$, it is possible to derive the radial component of its peculiar velocity, $u = cz - H_0 d$, where $c$ is the speed of light in vacuum, and $H_0$ is the Hubble constant. However, as distance moduli have Gaussian distributed errors, the peculiar velocities computed with this equation have non-Gaussian (skewed) distributed errors. To solve this problem, \cite{Watkins:2015aa} introduced a new estimator which results in Gaussian distributed errors on peculiar velocities:
\begin{equation}
u = cz \ln \left( \frac{cz}{H_0 d} \right).
\label{eq:vpec}
\end{equation}

Equation \ref{eq:vpec} will be used throughout this paper to derive radial peculiar velocities from observed distances.
 
This article will only focus on the CF3 distances catalog. Two radial peculiar velocity samples will be considered: the ungrouped sample and the grouped sample, containing galaxies and groups of galaxies respectively. Groups are frequently used by authors because they allow to reduce uncertainties with an $\sqrt{N}$ improvement on observed distances, and thus on radial peculiar velocities. These uncertainties are due to the virial motions of group members. However, the methodology presented in this article is valid for pairs of galaxies, and not for pairs of groups of galaxies. The CF3 grouped catalog is tested in this article since recent studies \citep[both introduced above in Section \ref{sec:intro}]{Ma:2015aa,Nusser:2017aa} made use of the grouped versions of the {\it Cosmicflows} catalogs to derive $f\sigma_8$. However, it will be seen in the discussion that using grouped data to constrain the growth rate leads to incoherent results.

\cite{Tully:2016aa} shows that the most consistent value of the Hubble constant with CF3 distances when computing radial peculiar velocities is $H_0 = 75 \pm 2$ km s$^{-1}$ Mpc$^{-1}$. This value is preferred as it minimizes the monopole term with CF3 distances and results in a tiny global radial infall and outflow in the peculiar velocity field. A larger value of $H_0$ would give a large overall radial infall towards the position of the observer, while choosing a smaller $H_0$ would yield a large radial outflow \citep[cf. Figure 21  in][]{Tully:2016aa}. 

For these reasons, in this article the value $H_0 = 75$ km s$^{-1}$ Mpc$^{-1}$ is used to compute radial peculiar velocities of CF3 galaxies and groups. We note that this high value of $H_0$ is consistent with other values of the Hubble constant measured in the local universe.

\subsection{Cosmicflows mock catalogs} 

A three-dimensional peculiar velocity field computed with the Constrained Realization (CR) methodology \citep{Hoffman:1991aa} is considered to construct mock catalogs. Using the CF3 grouped dataset and assuming a $\Lambda$CDM cosmological model ($\Omega_m=0.3$,  dark energy density parameter $\Omega_\Lambda=0.7$ and $H_0=70$ km s$^{-1}$ Mpc$^{-1}$), this velocity field is composed of a velocity field obtained with the Wiener-Filter technique \citep[WF]{Zaroubi:1995aa,Zaroubi:1999aa} and a random component derived from the Random Realizations method. The velocity field used in this article is reconstructed in a box 2000 Mpc wide, in Cartesian Supergalactic coordinates and centered on the Milky Way, with $128^3$ cells.
 
In order to test the various estimators of the peculiar velocity correlation function, mock catalogs are prepared as explained hereafter. Radial peculiar velocities, predicted by the three-dimensional velocity field of a CR, are assigned to galaxies or groups from the CF3 data. The mock catalogs are prepared with the following method. Considering galaxies and groups of the CF3 catalog, their predicted three-dimensional peculiar velocities are extracted from the CRs' peculiar velocity field at the redshift positions of the galaxies. The radial part of the peculiar velocity, which is the only observable component, is then derived from the three-dimensional velocity.

Throughout this article, parameters that are not fixed by the cosmology of the CR are set to their Planck 2015 values \citep[$\Omega_\Lambda = 0.69$, $\Omega_m =  0.31$, $\sigma_8 = 0.82$]{Planck-Collaboration:2016aa}.

%%%%%%%%%%%%%%%%%%%%%methode%%%%%%%%%%%%%%%%%%%%%%%%%%%%%

\section{Methodology}

\begin{figure}
\begin{center}
\includegraphics[width=0.3\textwidth]{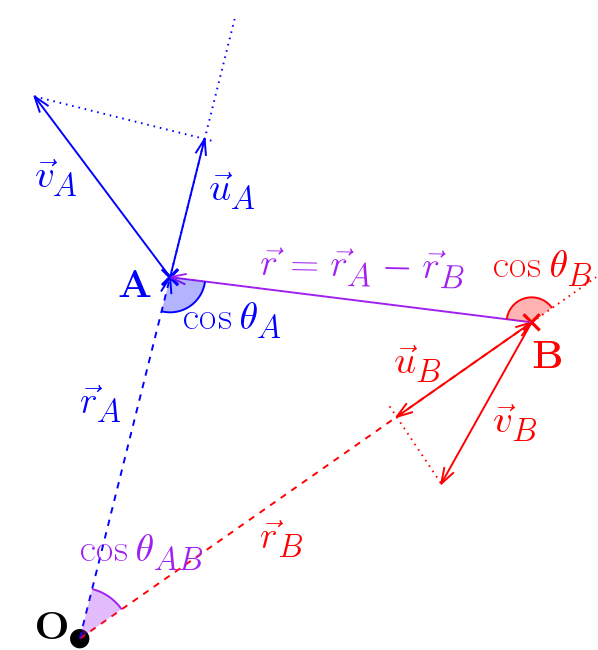}
\caption{Pair of galaxies considered throughout this article.}
\label{fig:galaxypair}
\end{center}
\end{figure}

We consider in this article a pair of galaxies $A$ and $B$ located at the positions $\vec{r}_A$ and $\vec{r}_B$ respectively. The spatial separation of these two galaxies is given by $\vec{r} = \vec{r}_A - \vec{r}_B$. Their peculiar velocities are $\vec{v}_A$ and $\vec{v}_B$ and their radial components are given by $\vec{u}_A = u_A \hat{\vec{r}}_A = \left(\vec{v}_A \cdot \hat{\vec{r}}_A\right) \hat{\vec{r}}_A$ and $\vec{u}_B = u_B\hat{\vec{r}}_B = \left(\vec{v}_B \cdot \hat{\vec{r}}_B\right) \hat{\vec{r}}_B$, where $\hat{\vec{r}}_{A,B}$ are the unit direction vectors of the galaxies. The cosines of the angles between the different directions are given by $\cos\theta_A = \hat{\vec{r}}_A \cdot \hat{\vec{r}}$, $\cos\theta_B = \hat{\vec{r}}_B \cdot \hat{\vec{r}}$ and $\cos\theta_{AB} = \hat{\vec{r}}_A \cdot \hat{\vec{r}}_B$. Figure \ref{fig:galaxypair} illustrates the geometry and quantities defined above.

To partially avoid the Malmquist bias, the galaxies (or groups) are located at their redshift positions, as positioning objects at their observed distance leads to much larger errors, especially for the most distant ones. 

\subsection{Mean pairwise velocity}

\subsubsection{Model}

The mean pairwise velocity $v_{12}$ was introduced for the first time in the context of the Bogolyubov-Born-Green-Kirkwood-Yvon theory \citep[BBGKY,][]{Yvon:1935aa,Bogoliubov:1946aa,Kirkwood:1946aa,Kirkwood:1947aa,Born:1946aa}. In this theory, the conservation equation of pairs of galaxies links the two-point correlation function $\xi(r)$ to the growth rate of large scale structures $f$ and the mean pairwise velocity $v_{12}\vec{r}/r$ \citep{Davis:1977aa,Peebles:1980aa}. For a pair of galaxies separated by a distance $r$, the mean pairwise velocity is given by \citep{Juszkiewicz:1998aa}:

\begin{equation}
v_{12} (r) = \left< \left( \vec{u_A} - \vec{u_B} \right) \cdot \hat{\vec{r}} \right>_\rho = \frac{\left<\left( \vec{u_A} - \vec{u_B} \right) (1 + \delta_A) (1+\delta_B)\right>}{1 + \xi(r)},
\end{equation}

\noindent where $\vec{u}_i$ and $\delta_i$ correspond respectively to the peculiar velocity and the density contrast at the location of the galaxies $i = A,B$, and $<...>_\rho$ specifies a pair-weighted average with $(1 + \delta_A) (1+\delta_B)(1 + \xi(r))^{-1}$ as the weighting factor.

In the non-linear regime, i.e for pairs of very close galaxies, $\xi(r)\gg 1$, and the solution of the pair conservation equation is $v_{12}(r) = -H_0 r$. In the case of the linear regime, i.e for large separation distances, $\xi(r) \ll 1$ and the solution of the conservation equation for $v_{12}$ is then given by the perturbative expansion of $\xi(r)$. In order to measure cosmological parameters such as the total matter density parameter $\Omega_m$, \cite{Juszkiewicz:1999aa} introduce a solution for $v_{12}$ valid in both regimes, linear and non-linear, by interpolating the linear and non-linear solutions:

\begin{equation}
v_{12}(r) \approx - \frac{2}{3} H r f \bar{\bar{\xi}}(r) \left[ 1 + \alpha \bar{\bar{\xi}}(r) \right],
\label{eq:modelv12}
\end{equation}

\noindent where $\bar{\bar{\xi}}(r) = \bar{\xi}(r) / [ 1+\xi(r) ]$ and $\bar{\xi}(r) = 3r^{-3} \int_0^r{ \xi(x) x^2 dx }$ is the two-point correlation function averaged in a sphere of radius $r$. The parameter $\alpha = 1.2 - 0.65 \gamma$ depends on the logarithmic slope of $\xi(r)$ denoted by the quantity $0<\gamma<3$ given by: 

\begin{equation}
\gamma = - \left. \frac{ d \ln \xi(r) }{ d \ln r } \right|_{\xi=1}.
\end{equation}

From the approximate solution for $v_{12}$ (equation \ref{eq:modelv12}), it is possible to recover the linear solution if $\xi \to 0$, and the solution valid in the non-linear regime if $x \to 0$. Equation \ref{eq:modelv12} has been tested and validated by \cite{Juszkiewicz:1999aa} on N-body simulations for $0.1 < \xi(r) < 1000$.

\subsubsection{Estimator}

\begin{figure*}
\begin{center}
\includegraphics[width=0.3\textwidth]{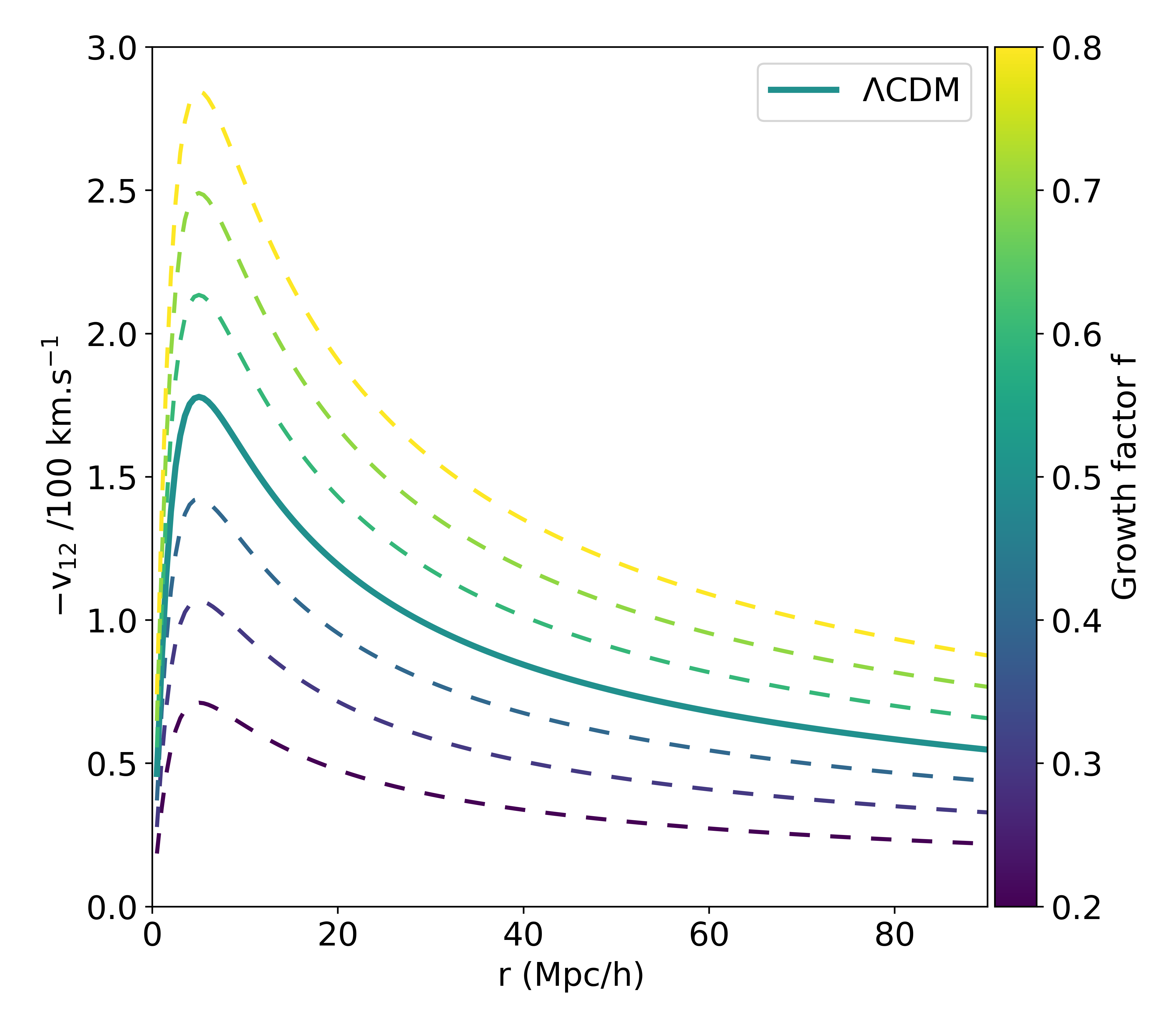}\includegraphics[width=0.6\textwidth]{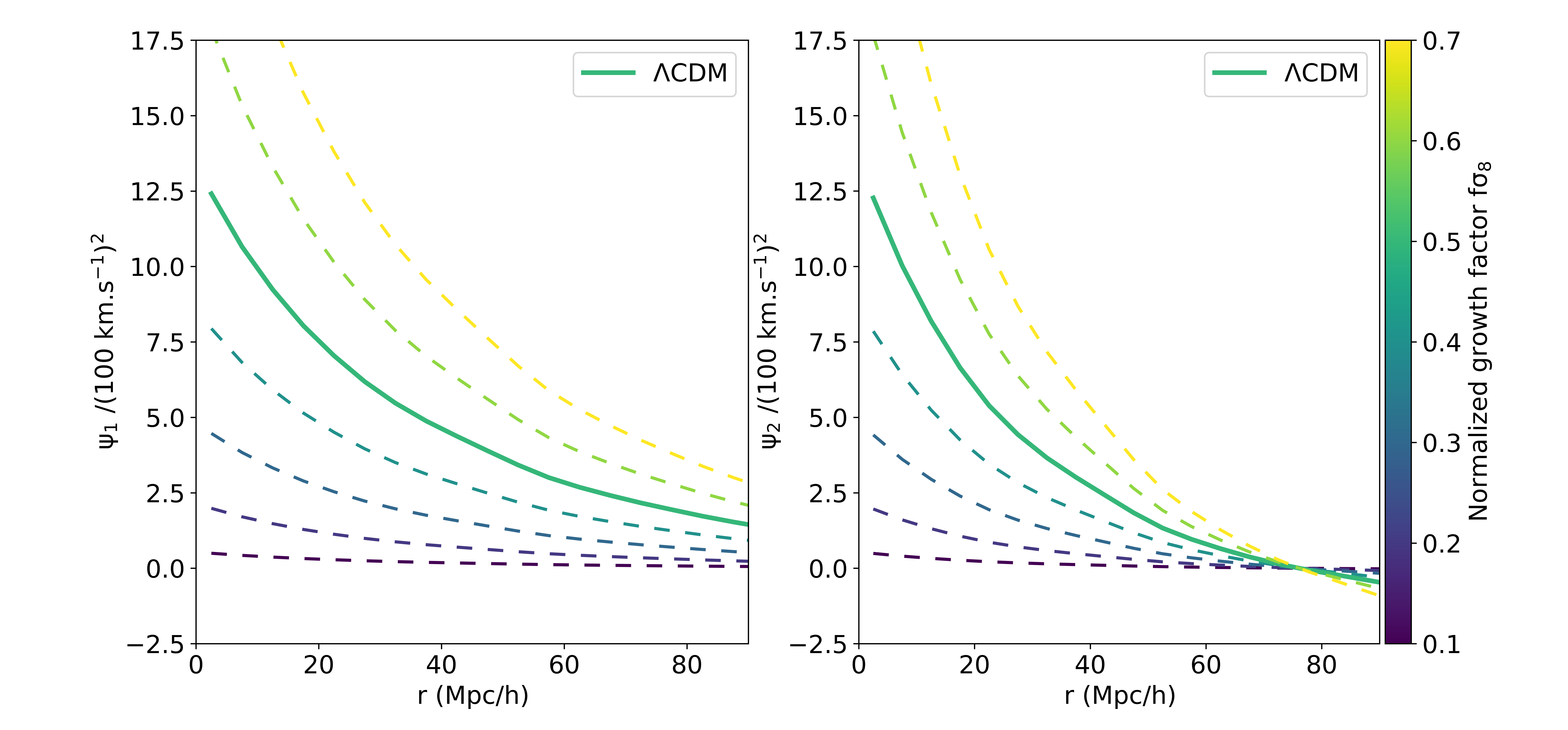}
\caption{$\Lambda$CDM models of the three usual peculiar velocity statistics $v_{12}$ (left), $\psi_1$ (middle) and $\psi_2$ (right) as a function of the pair separation $r$. The three statistics are computed for several values of $f\sigma_8$ (see colobar).}
\label{fig:models}
\end{center}
\end{figure*}

Equation \ref{eq:modelv12} shows that the amplitude of the mean pairwise velocity $v_{12}$ is related to the growth rate $f$, as shown in the left panel of Figure \ref{fig:models}. The statistic $v_{12}$ can therefore be used to constrain this parameter. However, observations give access only to the radial part of the peculiar velocities of galaxies. Therefore, one cannot use equation \ref{eq:modelv12} to compute $v_{12}$ directly from observed data. An estimator that can be used to compute $v_{12}(r)$ directly from observed radial peculiar velocities has to be considered. \cite{Ferreira:1999ab} introduced such an estimator to determine the mean pairwise velocity directly for observational catalogs of peculiar velocities:

\begin{equation}
v_{12}(r) = \frac{ 2 \sum{ (u_A - u_B) (\cos\theta_A + \cos\theta_B) } }{ \sum{ (\cos\theta_A + \cos\theta_B)^2 } },
\label{eq:estimatorv12}
\end{equation}

\noindent where the sums are computed for all pairs separated by a distance $r$.

A new estimator which relies on the transverse component of peculiar velocities (instead of the radial one) has been introduced by \cite{Yasini:2018aa}. This estimator will allow to analyze pairwise velocities derived from upcoming transverse peculiar velocity surveys such as Gaia \citep{Hall:2018aa}.

\subsection{Velocity correlation function}

In the linear regime, the two-point correlation tensor of a homogeneous and random velocity field $\vec{v}(\vec{r})$ is defined as \citep{Monin:1975aa,Strauss:1995aa}:

\begin{equation}
\Psi_{ij} (r) \equiv \left< v_i(\vec{r}_A) v_j(\vec{r}_B) \right> = \left[ \Psi_\parallel(r) - \Psi_\perp(r) \right] \hat{r}_i \hat{r}_j + \Psi_\perp(r) \delta_{ij},
\label{eq:correlationtensor}
\end{equation}

\noindent where $i$ and $j$ are the Cartesian coordinates. The quantities $\Psi_\parallel(r)$ and $\Psi_\perp(r)$ are the radial (i.e along $\vec{r}$) and transverse velocity correlation functions respectively. The spectral representations of $\Psi_\parallel(r)$ and $\Psi_\perp(r)$ are given by \citep{Gorski:1988aa}:

\begin{equation}
\Psi_\parallel(r)  = \frac{H_0^2 (f\sigma_8)^2}{2\pi^2} \int{ P_0(k) \left[ j_0(kr) - 2 \frac{j_1(kr)}{kr} \right] dk }
\end{equation}

\noindent and

\begin{equation}
\Psi_\perp(r)  = \frac{H_0^2 (f\sigma_8)^2}{2\pi^2} \int{ P_0(k) \frac{j_1(kr)}{kr} dk }
\end{equation}

\noindent where $j_0(x)$ and $j_1(x)$ are the spherical Bessel functions of the first kind:

\begin{equation}
j_0(x) = \frac{\sin x}{x}, \;\;\;\;\; j_1(x) = \frac{\sin x}{x^2} - \frac{\cos x}{x},
\end{equation}

\noindent and $P_0(k)$ is the non-normalized linear matter power spectrum measured today. In the rest of this article $P_0(k)$ is computed with CAMB in the Planck 2015 cosmology. 

The quantities $\Psi_\parallel(r)$ and $\Psi_\perp(r)$ both depend on the parameter $(f\sigma_8)^2$. These correlation functions can therefore be used to constraint the combined cosmological parameter $f\sigma_8$, defined as the normalized growth rate of large scale structures.

\subsubsection{Estimator}

\cite{Gorski:1989aa} introduced two velocity statistics, noted here as $\psi_1$ and $\psi_2$, which depend only on radial peculiar velocities. The statistic $\psi_1$ is defined as:

\begin{equation}
\psi_1(r) = \frac{ \sum{ \vec{u}_A \cdot \vec{u}_B }}{ \sum{ \left( \hat{\vec{r}}_A \cdot \hat{\vec{r}}_B \right)^2 } } = \frac{ \sum{u_A u_B \cos\theta_{AB}} }{ \sum{\cos^2\theta_{AB}} },
\label{eq:psi1}
\end{equation}

\noindent and $\psi_2$ as:

\begin{equation}
\psi_2(r) = \frac{ \sum{ \left( \vec{u}_A \cdot \hat{\vec{r}} \right) \left( \vec{u}_B \cdot \hat{\vec{r}} \right) } }{ \sum{ \left( \hat{\vec{r}}_A \cdot \hat{\vec{r}}_B \right) \left( \hat{\vec{r}}_A \cdot \hat{\vec{r}} \right) \left( \hat{\vec{r}}_B \cdot \hat{\vec{r}} \right) } } = \frac{ \sum{u_A u_B \cos\theta_A \cos\theta_B} }{ \sum{ \cos\theta_{AB}\cos\theta_A\cos\theta_B } },
\label{eq:psi2}
\end{equation}

\noindent The sums in equations \ref{eq:psi1} and \ref{eq:psi2} are performed over all pairs with fixed separation $r$. The denominators normalize the sums in order to preserve the norm of the velocity field.

\subsubsection{Model}

The correlation function of radial peculiar velocities can be derived from the two-point velocity correlation tensor: 

\begin{equation}
\left< u_{m}(\vec{r}_A) u_{n}(\vec{r}_B) \right> = \hat{r}_{Am}\hat{r}_{Bn} \Psi_{ij}(r)\hat{r}_{Ai}\hat{r}_{Bj},
\label{eq:radialcorr}
\end{equation}

\noindent where $i$, $j$, $m$ and $n$ are the Cartesian coordinates. Inserting equation \ref{eq:radialcorr} into equations \ref{eq:psi1} and \ref{eq:psi2}, the quantities $\psi_1(r)$ and $\psi_2(r)$ can be written as functions of $\Psi_\parallel(r)$ and $\Psi_\perp(r)$ \citep{Gorski:1989aa}:

\begin{equation}
\psi_{1,2} (r) = \mathcal{A}_{1,2} (r) \Psi_\parallel (r) + \left[ 1 - \mathcal{A}_{1,2}(r) \right] \Psi_\perp(r),
\label{eq:modelpsi12}
\end{equation}

\noindent where

\begin{equation}
\mathcal{A}_1(r) = \frac{ \sum{\cos\theta_A\cos\theta_B\cos\theta_{AB}} }{ \sum{\cos^2\theta_{AB}} },
\label{eq:c1}
\end{equation}

\noindent and

\begin{equation}
\mathcal{A}_2(r) = \frac{ \sum{\cos^2\theta_A\cos^2\theta_B} }{ \sum{\cos\theta_A\cos\theta_B\cos\theta_{AB}} }.
\label{eq:c2}
\end{equation}

The functions $\mathcal{A}_1(r)$ and $\mathcal{A}_2(r)$ contain information about the geometry of the sample, and measure the contributions of $\Psi_\parallel(r)$ and $\Psi_\perp(r)$ to the functions $\psi_1(r)$ and $\psi_2(r)$.

As one can see on the middle and right panels of Figure \ref{fig:models}, the amplitude of the two statistics $\psi_1$ and $\psi_2$ depends on the growth factor $f\sigma_8$, allowing to constrain this cosmological parameter. The higher $f\sigma_8$ is, the higher the amplitude of $\psi_1$ or $\psi_2$ is: a universe with a large $f\sigma_8$ appears more compact and peculiar velocities get larger. However, in the case of $\psi_2$, the curves corresponding to different $f\sigma_8$ get closer and closer as $r$ increases (see  right panel of Figure \ref{fig:models}). It is difficult to constrain cosmological models for separation distances higher than 60 Mpc/$h$. The statistic $\psi_2$ is therefore not robust enough to estimate the growth rate on peculiar velocity catalogs such as CF3, and especially on the upcoming big surveys. Therefore, this statistic will not be considered for the rest of this paper.

\subsection{Observational errors and cosmic variance}
\label{sec:errors}

Two kinds of uncertainties on peculiar velocity statistics are considered in this article: measurement (or observational) error and cosmic variance.

Observational errors on peculiar velocity statistics $v_{12}$, $\psi_1$ and $\psi_2$, i.e errors due to the uncertainty in distance measurement, are derived by Monte-Carlo synthetic realizations. These realizations are constructed by adding a random error to the radial peculiar velocity. The random error is extracted from a normal distribution with a standard deviation equal to the measurement error on the peculiar velocity. This measurement error is derived from the uncertainty on the distance (or distance modulus). In this article, 100 realizations have been computed for each sample (mocks and observed data).

In addition to measurements uncertainties, one ought to take into account the impact of cosmic variance when estimating cosmological parameters. However, computing uncertainties caused by cosmic variance cannot be done on a Constrained Realization of CF3, described in section \ref{sec:data}, which represents our local universe. Therefore, a $\Lambda$CDM dark matter only $N$-body simulation is considered in this paper in order to estimate the impact of the cosmic variance on peculiar velocity statistics and growth rate measurements. Two tests, whose results are presented later in this paper, on dark matter halos extracted from the MultiDark Planck 2 simulation \citep[MDPL2,][]{Prada:2012aa} of size 1 Gpc $h^{-1}$ are carried out. The underlying cosmology of the simulation is the Planck 2015 cosmology. Only mock galaxies with halo mass between $10^{11}$ M$_\odot$ and $10^{12}$ M$_\odot$ are taken into account to construct these mocks. Data samples for the tests are prepared as follows:

\begin{itemize}
\item as the CF3 catalog can (mostly) be contained in a sphere of a 250 Mpc $h^{-1}$ radius, one can place 8 of such independent spheres in a cube of side 1 Gpc $h^{-1}$. Therefore 8 CF3-like samples are generated for each octant cube of the simulation. The observer of each sample is placed at the center of its associated octant. Then radial components of peculiar velocities are extracted at the position of CF3 galaxies with respect to the observer. These mocks are completely independent from each other as they do not share any halos.
\item As only 8 samples is not high enough to get a robust result, a total of 100 more mocks are generated. Instead of positioning observers such that samples do not share halos, a total of 100 observers are placed at a random locations within the simulation box. Then radial components of peculiar velocities are extracted at the position of CF3 galaxies with respect to the observers. In this case, the samples are not independent as a single halo can belong to several spheres, so results will be correlated.
\end{itemize}

Results obtained from these tests are shown in section \ref{sec:results}.

As different bins of separation distances share the same galaxies (or groups of galaxies), errors between bins are correlated and thus the covariance matrix needs to be considered when fitting the measured statistics to extract the normalized growth rate $f\sigma_8$.

From these Monte-Carlo realizations or cosmic variance mocks, the covariance matrix $C$ between bins of separation distances of galaxy pairs can be computed. The covariance between bins $r_m$ and $r_n$ is computed as:
\begin{equation}
C_{mn} = \frac{1}{N_k-1} \sum_{k=1}^{N_k}{\left(S_k(r_m) - \mean{S_m}\right) \left(S_k(r_n) - \mean{S_n}\right)}
\end{equation}
where $S$ denotes the statistic considered and $\mean{S_{m,n}} = \frac{1}{N_k}\sum_{k=1}^{N_k}{S_k(r_{m,n})}$ is defined as the mean of all realizations for the bins $r_m$ and $r_n$ respectively and $N_k = 100$ is the number of mocks (realizations).

%%%%%%%%%%%%%%%%%%%%%resultats%%%%%%%%%%%%%%%%%%%%%%%%%%%%%

\section{Deriving the local growth rate}
\label{sec:results}

\subsection{Verification of estimators on mocks}

The normalized growth rate $f\sigma_8$ is estimated by fitting the theoretical models of the statistics $v_{12}$ and $\psi_1$, noted $S_\mathrm{mod}$ and computed with equation \ref{eq:modelv12} and \ref{eq:modelpsi12}, to the quantity $S_\mathrm{meas}$ measured with the radial estimators of the two statistics defined in equations \ref{eq:estimatorv12} and \ref{eq:psi1} respectively. The value of the growth rate $f\sigma_8$ is obtained by minimizing the following chi-square function:

\begin{equation}
\begin{split}
\chi^2 (f\sigma_8) = \sum_{i,j=0}^{N_\mathrm{bins}}&\Big[\Big(S_\mathrm{meas}(r_i) - S_\mathrm{mod}(r_i;f\sigma_8)\Big) C_{ij}^{-1} \\
 & \Big(S_\mathrm{meas}(r_j) - S_\mathrm{mod}(r_j;f\sigma_8)\Big) \Big]
\end{split}
\label{eq:chi2}
\end{equation}

\noindent The minimization is conducted with \texttt{MINUIT} \cite[Function Minimization and Error Analysis software,][]{James:1975aa}. The error on the fitted parameter is also given by \texttt{MINUIT} as the second derivative of the chi-square.

Peculiar velocity statistics $v_{12}$ and $\psi_1$ have been measured on mock and observed radial peculiar velocities. Throughout this article, statistics are computed out to a distance of 100 Mpc $h^{-1}$ in 20 equal bins of 5 Mpc $h^{-1}$. In all figures of this article displaying velocity statistics computed on mocks or observed data, scattered points are located at the middle of the bins.

\begin{figure*}
\begin{center}
\includegraphics[width=0.4\textwidth]{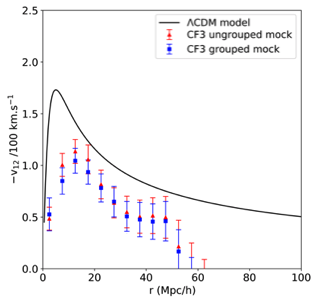}\includegraphics[width=0.4\textwidth]{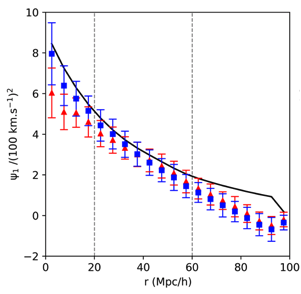}
\caption{The two peculiar velocity statistics $v_{12}$ (left) and $\psi_1$ (right) as a function of the pair separation $r$. The $\Lambda$CDM model ($\Omega_m=0.3$, $\gamma=0.55$, $\sigma_8=0.82$) is shown as a black solid line. Scattered points with error bars represent results obtained from mock peculiar velocities constructed from a $\Lambda$CDM constrained realization of CF3. The CF3 ungrouped and grouped mocks are shown as red triangles and blue squares respectively. Vertical dashed lines show the region where $f\sigma_8$ can be fitted, see the limitations described in the text.}
\label{fig:mocks}
\end{center}
\end{figure*}

The statistics $v_{12}$ and $\psi_1$ have been tested and validated by the authors who introduced them. They allow one to recover the underlying cosmology from a homogeneous and spherical universe. But CF3 is very sparse and asymmetrical. Before constraining the growth rate, one must check if the spatial distribution of the CF3 catalog alone inhibits such statistics from accurately recovering the underlying cosmology. This is done on the 100 CF3 mocks generated from a Constrained Realization as described in section \ref{sec:data}.

Figure \ref{fig:mocks} shows as solid lines the $\Lambda$CDM models for galaxies in the CF3 ungrouped (red) and CF3 grouped (blue) mocks. For the statistic $v_{12}$, errors bars of the two mocks do not include the $\Lambda$CDM model. This means that due to the unique CF3 geometry and selection function, this estimator can not recover the underlying $\Lambda$CDM cosmology. Therefore we stress that it cannot be used to estimate to local growth rate with the real CF3 catalog, and will not be considered in the analysis that follows. Also, this explains why \cite{Ma:2015aa} obtained incoherent values for $\Omega_m$ and $\sigma_8$ by applying $v_{12}$ on the CF2 dataset, whose footprint is similarly inhomogeneous. For the estimator $\psi_1$, the amplitude of the CF3 mocks (red and blue dots with error bars for the ungrouped and grouped samples respectively) is slightly lower but error bars are consistent with the models up to 60 Mpc $h^{-1}$. This shows that the geometry and the sparseness of the current survey prevent from deriving any growth rate for separation distances larger than 60 Mpc $h^{-1}$. 

Furthermore, one can see in Figure \ref{fig:mocks} that for both statistics the overall differences between the CF3 grouped and ungrouped samples are very small. This shows that non-linearity does not have any impact on the $\psi_1$ estimator except for small separations bins which contain galaxies close to each other (i.e within clusters). When fitting $f\sigma_8$ to this statistic, the effect of non-linearity will not be taken into account: bins corresponding to separations lower than 20 Mpc $h^{-1}$ will be omitted. 

Tests on mocks reported in Figure \ref{fig:mocks} show that the underlying value of $f\sigma_8$ in the CR is recovered by the estimator $\psi_1$ in an interval of robustness 20 -- 60 Mpc/$h$. However, considering the depth of the CF3 catalog and the size of the constrained realization, the measured value of the growth rate with CF3 datasets gives only its local value. Due to cosmic variance, this local value may not represent the global value of the growth rate of large scale structures of the entire universe.
 
 \begin{figure}
\begin{center}
\includegraphics[width=0.4\textwidth]{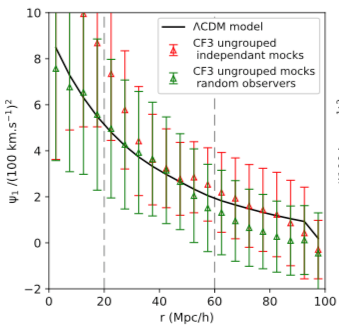}
\caption{Peculiar velocity statistic $\psi_1$ as a function of the pair separation $r$ computed in the large MDPL2 simulation box. The simulation's underlying $\Lambda$CDM model is shown as a black solid line. Red scattered triangles with error bars represent results obtained from the 8 CF3-like independent mocks. Green scattered triangles with error bars represent results obtained from the 100 CF3-like mocks constructed with randomly positioned observers. Vertical dashed lines show the region where where $f\sigma_8$ can be fitted, see the limitations described in the text.}
\label{fig:cosmicvar}
\end{center}
\end{figure}
 
Figure \ref{fig:cosmicvar} shows peculiar velocity statistics obtained with the 8 independent mocks extracted from the MDPL2 simulation (see section \ref{sec:errors}) as red triangles with error bars. The underlying cosmology of the simulation, shown by the black line, is well recovered by the estimator $\psi_1(r)$. Results obtained with the 100 CF3-like mocks generated considering random observers in the MDPL simulation is shown in Figure \ref{fig:cosmicvar} as green triangles with error bars. The underlying cosmology of the simulation is once again well recovered. This shows that despite its depth, the CF3 catalog may allow measuring the growth rate of large scale structures. Nevertheless, errors bars due to cosmic variance and the extent of the sample are large and cannot be ignored. They must be taken into account when constraining $f\sigma_8$ with CF3 data.

\begin{figure}
\begin{center}
\includegraphics[width=0.5\textwidth]{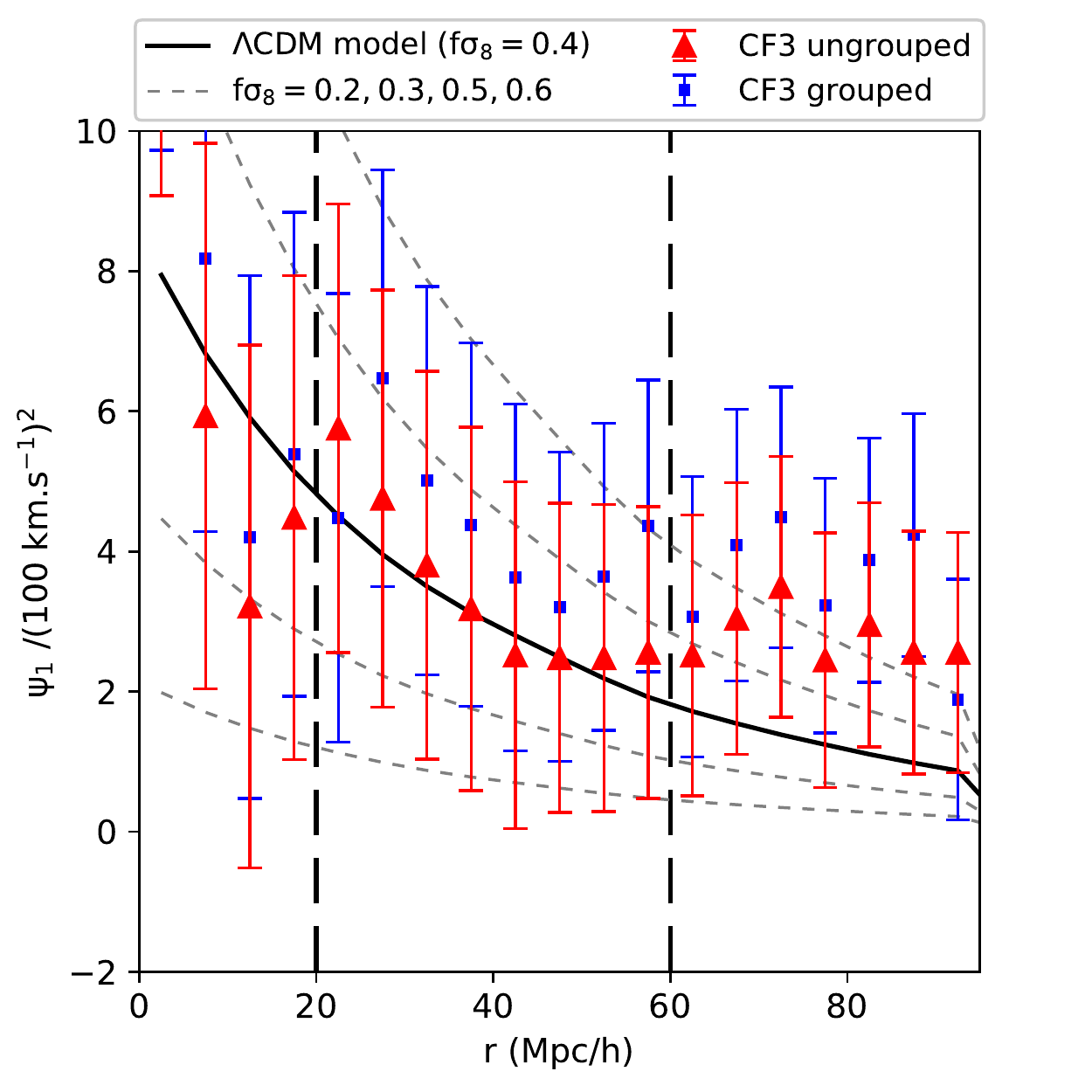}
\caption{Observed peculiar velocity statistic $\psi_1$ as a function of the pair separation $r$. The $\Lambda$CDM model, $f\sigma_8=0.4$, predicted for CF3 galaxies is shown as a black solid line. Peculiar velocity statistics predicted for other cosmological models, $f\sigma_8=0.2,0.3,0.5,0.6$, are shown as black dashed and dotted lines. Scattered points with error bars represent results obtained from observed peculiar velocities. The ungrouped sample of CF3 is shown as red triangles and the grouped CF3 sample is shown as blue squares. Vertical dashed lines show the region where $f\sigma_8$ can be fitted, see the limitations described in the text.}
\label{fig:observed}
\end{center}
\end{figure}

\subsection{Estimation of $f\sigma_8$ on data}

Figure \ref{fig:observed} shows the results of the computation of $\psi_1$ on the CF3 observations, for the ungrouped and grouped samples represented respectively as red triangles and blue squares. Errors bars include both, observational and cosmic variance uncertainties. Predictions of statistic $\psi_1$ for different cosmological models are shown as solid ($\Lambda$CDM, $f\sigma_8=0.4$) and dotted ($f\sigma_8=0.2,0.3,0.5,0.6$) lines. The two estimators obtained for CF3 peculiar velocities are consistent with $\Lambda$CDM within their interval of robustness, as predicted by tests on mocks shown in Figure \ref{fig:mocks}. Besides, one can observe that results obtained from CF3 groups have a slightly larger amplitude than results derived from CF3 galaxies.

For completeness, $\psi_1$ has also been applied to the previous CF2 catalog (not shown). The results are similar to CF3 and also in agreement with $\Lambda$CDM.

\begin{table}
\caption{Constraints of the normalized growth rate $f\sigma_8$ obtained from measurements of the peculiar velocity statistic $\psi_1$, within its interval of robustness as described in the text, on the {\it Cosmicflows} datasets of observed peculiar velocities. The two last columns $\Delta f\sigma_8^\mathrm{obs}$ and $\Delta f\sigma_8^\mathrm{cosmic}$ correspond to observational and cosmic variance uncertainties on the constrained parameter respectively.}
\begin{center}
\begin{tabular}{|c|c|c|c|c|c|}
\hline
CF3 sample --- statistic & $f\sigma_8$ & & $\Delta f\sigma_8^\mathrm{obs}$ & & $\Delta f\sigma_8^\mathrm{cosmic}$ \\
\hline
{\bf CF3 ungrouped} & $\boldsymbol{0.43}$ & $\boldsymbol{\pm}$&$\boldsymbol{0.03}$ & $\boldsymbol{\pm}$&$\boldsymbol{0.11}$ \\
\hline
%CF3 ungrouped --- $\psi_2$ & $0.51$ & $\pm$&$0.03$ & $\pm$&$0.12$ \\
CF3 grouped & $0.45$ & $\pm$&$0.04$ & $\pm$&$0.11$  \\
%CF3 grouped --- $\psi_2$ & $0.52$ & $\pm$&$0.04$ & $\pm$&$0.12$  \\
\hline
\end{tabular}
\end{center}
\label{tab:fits}
\end{table}%

The data in Figure \ref{fig:observed} are fitted by minimizing the $\chi^2$ defined in equation \ref{eq:chi2}. The fits are done within the interval [20, 60] Mpc $h^{-1}$ of pair separation distances. The results are displayed in Table \ref{tab:fits} which shows the values of $f\sigma_8$ fitted directly from observational data. Both observational and cosmic variance uncertainties on the constrained growth rate are reported. 

\begin{figure}
\begin{center}
\includegraphics[width=0.5\textwidth]{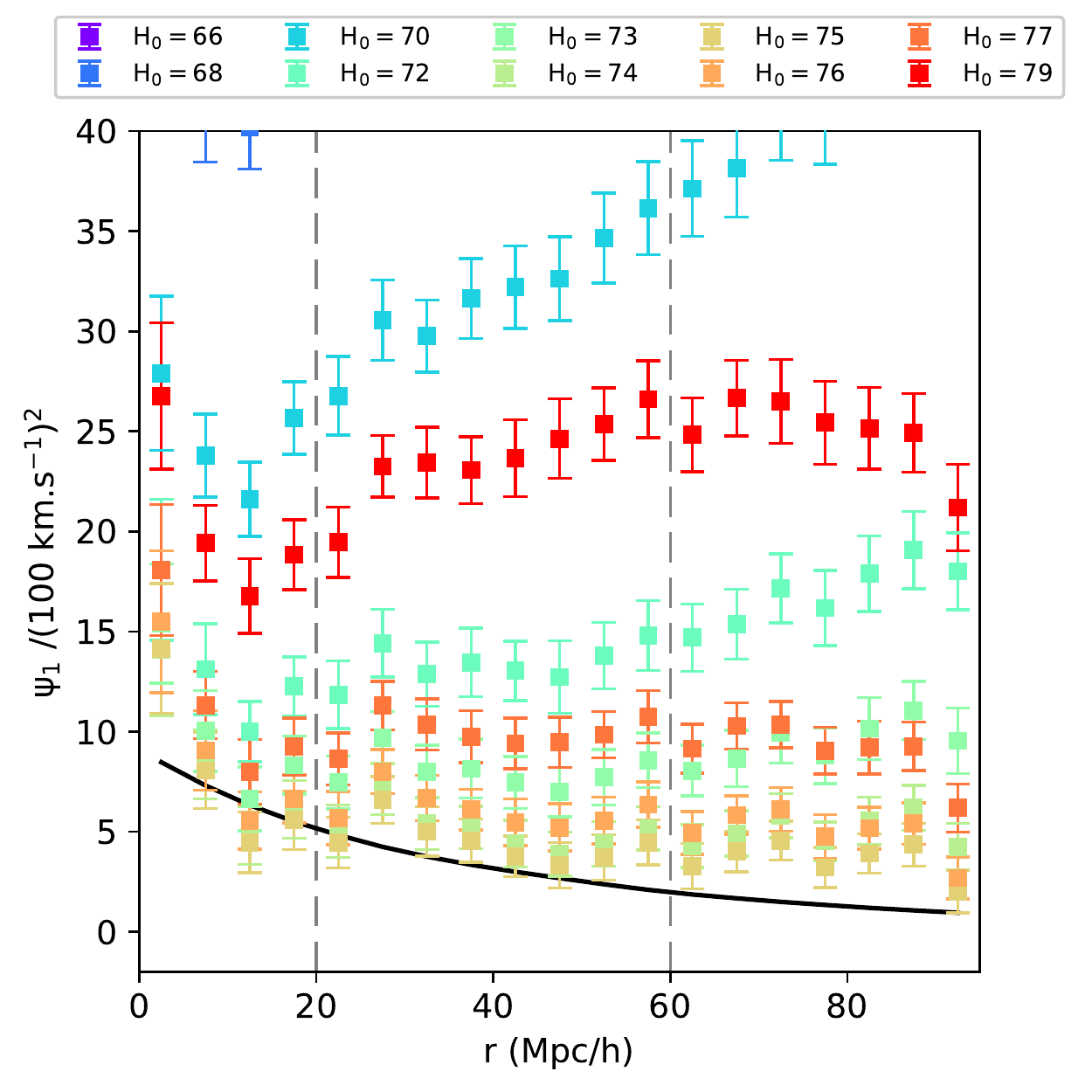}
\caption{Observed peculiar velocity statistic $\psi_1$ as a function of the pair separation $r$. The $\Lambda$CDM model, $f\sigma_8=0.4$, predicted for CF3 galaxies is shown as a black solid line. Scattered points with error bars represent results obtained from observed peculiar velocities of the CF3 galaxies, derived by considering different values of the Hubble constant $H_0=[66,68,70,72,73,74,75,76,77,79]$ km/s/Mpc.}
\label{fig:H0}
\end{center}
\end{figure}

Results shown in Figure \ref{fig:observed} and Table \ref{tab:fits} are obtained using the radial peculiar velocities computed from CF3 distances with equation \ref{eq:vpec} and $H_0 = 75$ km s$^{-1}$ Mpc$^{-1}$. However, \cite{Tully:2016aa} suggested that the range $H_0 = 75 \pm 2$ km s$^{-1}$ Mpc$^{-1}$ gives a reasonable value for the monopole flow. This uncertainty on the Hubble constant may affect the uncertainty on the growth rate measurements, as the amplitude of $\psi_1$ depends on the chosen value of $H_0$ when deriving radial peculiar velocities, as shown in Figure \ref{fig:H0}. Therefore, we considered other values of $H_0$ within the error range given in \cite{Tully:2016aa}. Taking $H_0 = 73$ km s$^{-1}$ Mpc$^{-1}$ to compute peculiar velocity of CF3 galaxies gives $f\sigma_8 = 0.58 \pm 0.03$, and taking $H_0 = 77$ km s$^{-1}$ Mpc$^{-1}$ gives $f\sigma_8 = 0.54 \pm 0.03$.

\cite{Wang:2018aa} show a similar study in the Fig7 of their paper. They published error bars that are twice smaller than the ones we derived for $\psi_1$ (see Fig8 of \cite{Wang:2018aa}).

\begin{figure*}
\begin{center}
\includegraphics[width=\textwidth]{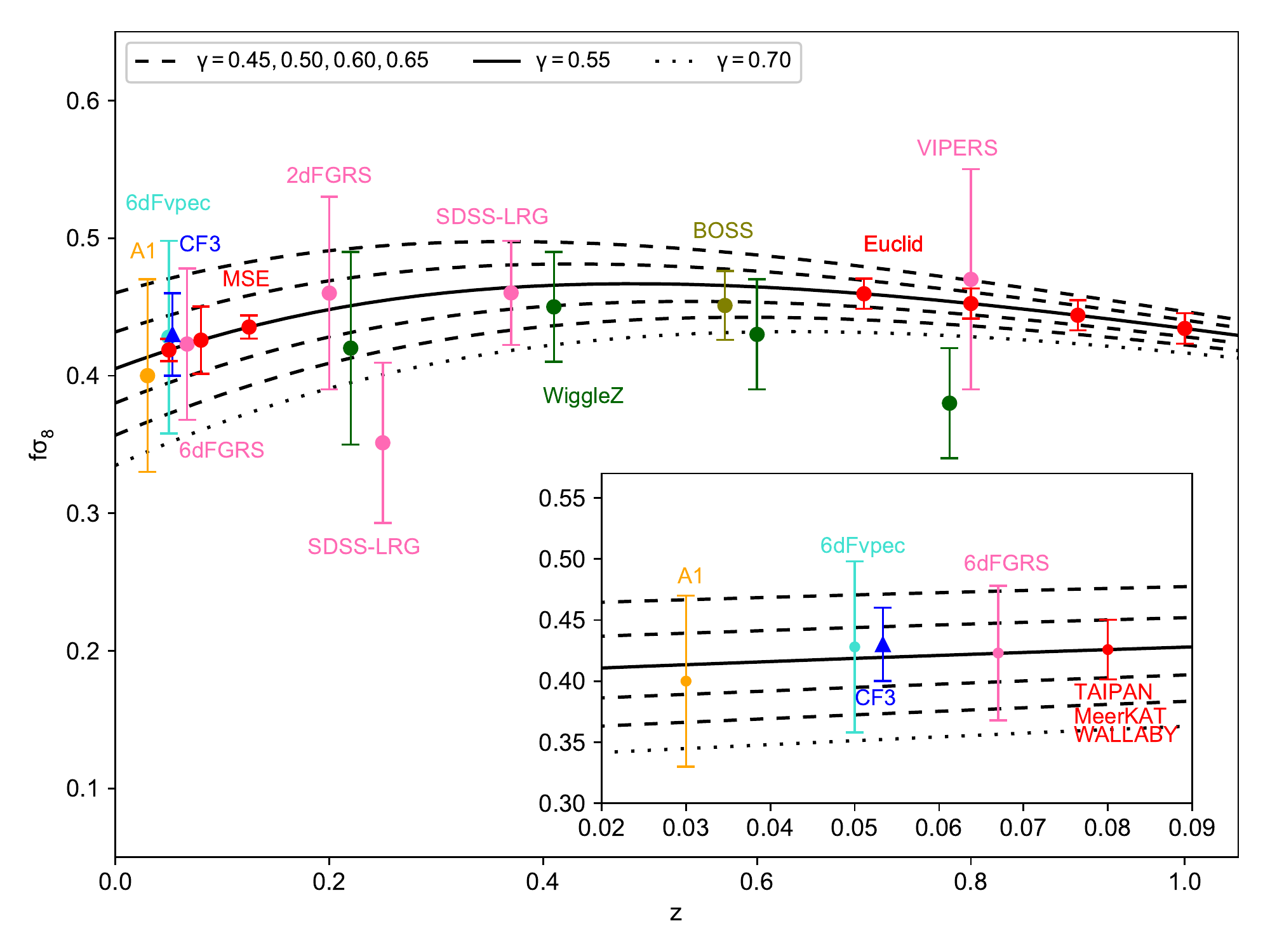}
\caption{Normalized growth rate $f\sigma_8$ as a function of the redshift $z$. The $\Lambda$CDM model ($\gamma=0.55$) is represented by a black solid line. Dotted and dashed black lines correspond to other cosmological models. Results taken from literature and displayed in Table \ref{tab:literature} are shown by scattered points with error bars. Colors depend on the cosmological probe used to constraint the growth rate: SN Ia in yellow, galaxy peculiar velocities in light blue, Baryon Acoustic Oscillations in green and Redshift Space Distortions in pink. Growth rate measurements predictions for future cosmological surveys are shown in red. Result of this article obtained from observed peculiar velocities of CF3  galaxies is represented by a blue point with error bars. Errors bars include observational uncertainties only.}
\label{fig:fs8z}
\end{center}
\end{figure*}

Figure \ref{fig:fs8z} shows the normalized growth rate $f\sigma_8$ as a function of redshift $z$. The black solid line corresponds to the $\Lambda$CDM model ($\gamma = 0.55$). Other cosmological models (modified gravity, $\gamma \neq 0.55$) are represented by dotted and dashed black lines. Scattered points with error bars show results taken from literature summarized in Table \ref{tab:literature}. Colors depend on the cosmological probe used to constraint the growth rate: SN Ia in yellow \citep{Turnbull:2012aa}, galaxy peculiar velocities in blue \citep{Johnson:2014aa}, Baryon Acoustic Oscillations in green \citep{Blake:2011aa,Reid:2012aa} and Redshift Space Distortions in pink \citep{Hawkins:2003aa,Samushia:2012aa,Beutler:2012aa,de-la-Torre:2013aa}. Growth rate measurements predictions for future cosmological surveys are shown in red, with a predicted error bar of 10 per cent \citep{McConnachie:2016aa,Amendola:2016aa,da-Cunha:2017aa}.

Local growth rate constraint obtained from radial peculiar velocities of CF3 galaxies is shown in Figure \ref{fig:fs8z} as a blue point with error bars. Errors bars include the observational uncertainties only, in coherence with other data points.

\begin{table*}
\caption{Measurements of normalized growth rate $f\sigma_8$ at redshift $z$ taken from literature. These constraints are obtained by using various cosmological probes: peculiar velocities (Vpec), Type Ia Supernovae (SN Ia), Redshift Space Distortions (RSD) and Baryon Acoustic Oscillations (BAO).}
\begin{center}
\begin{tabular}{|c|c|c|c|c|}
 & Redshift $z$ & Normalized growth rate $f\sigma_8$ & Publication & Cosmological probe \\
\hline
6dFGS & $0.05$ & $0.428^{+0.079}_{-0.068}$ &\cite{Johnson:2014aa}& Vpec \\
\hline
A1 & $0.03$ & $0.40\pm0.07$ &\cite{Turnbull:2012aa}& SN Ia \\
\hline
2dFGRS & $0.20$ & $0.46\pm0.07$ &\cite{Hawkins:2003aa}& RSD \\
\hline
6dFGRS & $0.067$ & $0.423\pm0.055$ &\cite{Beutler:2012aa}& RSD \\
\hline
WiggleZ & $0.22$ & $0.42\pm0.07$ &\cite{Blake:2011aa}& BAO\\
 & $0.41$ & $0.45\pm0.04$ && \\
 & $0.60$ & $0.43\pm0.04$ && \\
 & $0.78$ & $0.38\pm0.04$ && \\
\hline
SDSS-LRG & $0.25$ & $0.3512\pm0.0583$ &\cite{Samushia:2012aa}& RSD \\
 & $0.37$ & $0.4602\pm0.0378$ && \\
\hline
BOSS & $0.57$ & $0.451\pm0.025$ &\cite{Reid:2012aa}& BAO\\
\hline
VIPERS & $0.80$ & $0.47\pm0.08$ &\cite{de-la-Torre:2013aa}& RSD \\
\hline
\end{tabular}
\end{center}
\label{tab:literature}
\end{table*}%

%%%%%%%%%%%%%%%%%%%%%conclusion%%%%%%%%%%%%%%%%%%%%%%%%%%%%%

\section{Conclusion}

This article presents a measurement of the local value of $f\sigma_8$ in the nearby universe by means of a peculiar velocity survey: {\it Cosmicflows-3}. We obtained a measurement of $f \sigma_8 = 0.43 \left( \pm 0.03 \right)_\mathrm{obs} \left( \pm 0.11 \right)_\mathrm{cosmic}$ out to $z=0.05$. Uncertainties correspond to observational "$\mathrm{obs}$" and cosmic variance "$\mathrm{cosmic}$" uncertainties respectively. This is in complete agreement with the measurement made by \cite{Wang:2018aa}, finding $f\sigma_8 = 0.488$.

Currently, as seen in Figure \ref{fig:fs8z} some local cosmological probes like BAO and RSD do not have error bars that could constrain the growth rate of large scale structures. Even large redshift surveys such as SDSS, VIPERS or WiggleZ have a too large observational error budget to discriminate a cosmological model. In contrast, one can see that using only a few thousands of peculiar velocities of galaxies as cosmological probes to measure $f\sigma_8$ allows to restrict the range of cosmological models. However, due to the unique geometry of the CF3 survey as well as its sparseness, we have found that we are unable to use the full arsenal of velocity field statistics to probe $f\sigma_8$. We look forward to increased coverage to improve on this situation.

Upcoming redshift and peculiar velocity surveys (such as TAIPAN, MeerKAT, WALLABY, Euclid and MSE) are expected to improve uncertainties in a large range of redshifts from the very local universe up to $z=1.7$. Hopefully this will allow to constrain the results also in cosmic variance allowing finally to discriminate a single cosmological model using galaxy 2-point statistics of peculiar velocities.

%%%%%%%%%%%%%%%%%%%%%remerciements%%%%%%%%%%%%%%%%%%%%%%%%%%%%

\section*{Acknowledgments}

Noam Libeskind, Carlo Schimd, Romain Graziani, Yannick Copin and Mickael Rigault are gratefully thanked for scientific discussions.
Support has been provided by the Institut Universitaire de France and the CNES. 

\noindent The CosmoSim database used in this article is a service by the Leibniz-Institute for Astrophysics Potsdam (AIP).
The MultiDark database was developed in cooperation with the Spanish MultiDark Consolider Project CSD2009-00064.

%%%%%%%%%%%%%%%%%%%%biblio%%%%%%%%%%%%%%%%%%

\bibliographystyle{mnras}
\bibliography{biblio} 

%%%%%%%%%%%%%%%%%%%%%%%%%%%%%%%%%%%%%%%%%%%%%%%%%%

\bsp	% typesetting comment
\label{lastpage}
\end{document}